%% file: epl995.tex
\documentstyle[epsf]{europhys}

\input euromacr

\newcommand{\cac} {\varphi_{\rm cac}}
\newcommand{\cmc} {\varphi_{\rm cmc}}
\newcommand{\eg} {{\it e.g., }}
\newcommand{\ie} {{\it i.e., }}
\newcommand{\kB} {k_{\rm B}}
\newcommand{\vecr} {{\bf r}}

\begin{document}

\euro{}{}{}{}

\Date{}

\shorttitle
{H. DIAMANT \etal POLYMER-SURFACTANT SELF-ASSEMBLY}

\title
{Onset of self-assembly in polymer-surfactant systems}

\author{H. Diamant \And D. Andelman}

\institute
{School of Physics and Astronomy\\
 Raymond and Beverly Sackler Faculty of Exact Sciences\\
 Tel Aviv University, 69978 Tel Aviv, Israel}

\rec{}{}

\pacs{
\Pacs{61}{25Hq}{Macromolecular and polymer solutions, polymer melts, 
    swelling}
\Pacs{61}{41+e}{Polymers, elastomers, and plastics}
\Pacs{82}{65Dp}{Thermodynamics of surfaces and interfaces}
      }

\maketitle

\begin{abstract}

The onset of self-assembly in a dilute
aqueous solution containing a flexible polymer and
surfactant is theoretically studied.
Focusing on the effect of the surfactant on polymer
conformation and using a conjecture of 
partial collapse of the polymer at the onset of 
self-assembly, we obtain results which
agree with known experimental observations:
(i) polymer--surfactant self-assembly 
always starts at a lower concentration ({\em cac}) than
the one required for surfactant--surfactant self-assembly 
({\em cmc});
(ii) in charged systems the {\em cac} increases with 
salt concentration and is almost independent of
polymer charge;
(iii) in weakly interacting systems the {\em cac} remains
roughly proportional to the {\em cmc} over a wide range of
{\em cmc}  values.
The special case of amphiphilic side-chain polymers 
strongly supports our basic conjecture.
A similarity is found between the partial collapse induced 
by the surfactant and general results concerning the effect 
of impurities on critical phenomena.

\end{abstract}

Aqueous solutions containing polymers and surfactants
have been the subject of extensive research in the
past few decades \cite{review_general1,review_general2}.
The possibility to achieve polymer--surfactant aggregation 
at very low surfactant concentration offers a delicate 
control over the properties of the mixture, a feature being 
used in numerous applications \cite{review_application}.

The joint self-assembly of polymers and surfactants
usually starts at a well-defined surfactant concentration,
the {`critical aggregation concentration'} ({\em cac}).
One of the most consistent experimental observations in
polymer--surfactant systems is that the {\em cac} 
is found to be lower 
than the `critical micellar concentration' ({\em cmc}) of 
the polymer-free surfactant solution. 
Consequently, polymer--surfactant systems are commonly 
divided into two categories \cite{review_Goddard}:
(i) polyelectrolyte and oppositely charged ionic surfactant,
where the {\em cac} can become several orders of magnitude
lower than the {\em cmc};
(ii) neutral polymer and ionic surfactant,
where the {\em cac} is lower than, but comparable to the {\em cmc}.
Less common are systems containing a polyelectrolyte
and a non-ionic surfactant \cite{pe_nonionic},
which can be included in the second category as their {\em cac} is
comparable to the {\em cmc}.
Systems where both species are neutral exhibit a very
weak effect \cite{review_Lindman,Feitosa96}.

Several theories have been suggested for polymer--surfactant
aggregation \cite{micelle_models}-\cite{Blankschtein}.
They include various generalisations of micellisation
theories \cite{micelle} in the presence of a polymer.
However, most of those models do not explicitly treat
intra-chain features of the polymer.
This approach may be justified for rigid polymers such as
DNA or strong polyelectrolytes in the absence of salt. 
It is somewhat more questionable in view 
of the strong conformational changes observed in flexible
polymers upon self-assembly 
\cite{review_Lindman}.

This Letter takes a different point of view.
Instead of examining the effect of added polymer on
surfactant micellisation, we rather focus
on the effect of the surfactant on the statistics of a 
flexible polymer below and at the onset of self-assembly.
Details of the joint self-assembly above the
{\em cac} (\eg morphology of aggregates and their 
arrangement along the chain) probably depend on various 
molecular parameters. 
We suggest, however, that the {\em onset} of 
self-assembly in such systems may be accounted for by
more general considerations related to dilute mixtures
of polymers and smaller, interacting molecules.
We conjecture that in a flexible polymer--surfactant system
the {\em cac} is associated with a considerable change in
polymer statistics, \ie local instability of the polymer
chain.
This description is reminiscent of de Gennes' and
Brochard's treatment of a polymer in a binary mixture of 
good solvents close to its critical point \cite{deGennes}.
Similar to the latter scenario, the polymer is predicted
to undergo {\it partial collapse} \cite{deGennes}
at the {\em cac}, 
which marks the onset of aggregation.
This approach allows us to distinguish and explain 
some common, `universal' features in the vast 
experimental literature which has accumulated on 
polymer--surfactant systems. (See, \eg ref.
\cite{Blankschtein} for a more microscopic approach.)

The free energy of the polymer solution is assumed
to be characterised by a single interaction parameter
(2nd virial coefficient). Thus, the theory is applicable
to a dilute as well as semi-dilute polymer regime.
The theory is restricted, however, to the onset of binding 
({\em cac}) and its vicinity.
Issues of morphology, phase behaviour and rheology, 
especially in more concentrated polymer--surfactant systems, 
are very interesting and important \cite{review_Lindman},
but lie outside the scope of the current Letter.

Consider a dilute solution of polymer and 
surfactant whose local concentrations are denoted by
$c$ and $\varphi$, respectively. 
The polymer is assumed to be flexible and in a good solvent.
The surfactant is both below its {\em cmc} and {\em cac}.
The free energy density can be divided into three terms
accounting for the polymer contribution, the surfactant
one, and the polymer--surfactant interaction,
\begin{equation}
  f(c,\varphi) = f_{\rm p}(c) + f_{\rm s}(\varphi)
  + f_{\rm ps}(c,\varphi).
\end{equation}
Since the concentrations of both species are low and we 
are interested only in the onset of binding, 
we restrict to two-body interactions between monomers
and surfactants. Thus,
the leading quadratic
term in the expansion of $f_{\rm ps}(c,\varphi)$ is sufficient,
$
  f_{\rm ps}(c,\varphi) = -wc\varphi
$,
where $w\equiv -\partial^2 f_{\rm ps}/\partial c\partial\varphi$
is a parameter characterising the attraction strength.

In the absence of polymer the surfactant concentration 
has a homogeneous value, $\varphi\equiv\varphi_{\rm b}$, 
corresponding to the
minimum of the grand-canonical free energy of the surfactant,
$f_{\rm s}(\varphi)=h(\varphi)-\mu\varphi$, 
where $h$ denotes the canonical free energy and
$\mu$ is the surfactant chemical potential.
Consider a small perturbation in local surfactant
concentration, 
$\varphi=\varphi_{\rm b}+\delta\varphi$.
Since the solution is both below its {\em cac} and {\em cmc},
$f$ can be expanded in small $\delta\varphi$ to yield
\begin{equation}
  f = f_{\rm p}(c) + f_{\rm s}(\varphi_{\rm b}) - wc\varphi_{\rm b}
  - wc\delta\varphi
  + \frac12 f_{\rm s}^{''}(\varphi_{\rm b}) \delta\varphi^2,
\label{perturbed_f}
\end{equation}
where $f_{\rm s}^{''}(\varphi)\equiv
\partial^2 f_{\rm s}/\partial\varphi^2$.
We identify the {\em cmc} as the value of $\varphi$
at which, for $c=0$, the surfactant solution becomes unstable to
small perturbations, \ie
$
  f_{\rm s}^{''}(\cmc) = 0
$.
This simplification neglects various surfactant features,
assuming that surfactants can be considered for $\varphi<\cac$
merely as interacting small molecules, and their 
specific features be incorporated in the phenomenological
parameter $\cmc$.

Let $F(x)$ be a dimensionless function such that
$\varphi=\cmc F(x)$ solves the equation
$
  \cmc f_{\rm s}^{''}(\varphi) = 1/x
$.
Obviously, for $x\rightarrow\infty$ the solution is $\varphi=\cmc$,
\ie $F(x\gg 1)\simeq 1$.
In the other limit, $x\rightarrow 0$ and 
$f_{\rm s}^{''}(\varphi)\rightarrow\infty$, 
the solution for $\varphi$ must tend to zero independently 
of $\cmc$.
Hence, $F(x\ll 1)\sim x$.
This asymptotic behaviour is also obtained by
calculating explicit expressions for $F(x)$
in more specific models \cite{future}.
[For example, taking the simplest expression,
$f_{\rm s}=\varphi(\ln\varphi - 1) -  u\varphi^2/2$,
gives $F(x)=(1+1/x)^{-1}$.]

In the presence of the polymer, 
minimisation of eq.~(\ref{perturbed_f}) with respect to 
$\delta\varphi$ gives
$
  \delta\varphi = [w / f_{\rm s}^{''}(\varphi_{\rm b})] c
$,
and
$
  f = f_{\rm s}(\varphi_{\rm b}) + f_{\rm p}(c)
  - wc\varphi_{\rm b} 
  - w^2/[2f_{\rm s}^{''}(\varphi_{\rm b})] c^2
$.
Thus, the interaction with the surfactant leads to an 
effective reduction in the 2nd virial 
coefficient of the polymer \cite{monolayer},
\begin{equation}
  v_{\rm eff} = v - v_{\rm ps}; \ \ \ 
  v_{\rm ps} \equiv w^2/f_{\rm s}^{''}(\varphi_{\rm b}),
\label{veff}
\end{equation}
where $v$ is the bare 2nd virial coefficient.
The chain becomes unstable when $v_{\rm eff}=0$.
At this point the local polymer concentration, $c$, 
is expected to increase significantly
(due to contraction of chain conformation),
leading to a sharp increase in $\delta\varphi$ as well.
We identify this instability, therefore, as the
{\em cac}.
Using eq.~(\ref{veff}) and the function $F(x)$ defined above,
the following scaling expression for the {\em cac} is found:
\begin{equation}
  \cac = \cmc F[v/(w^2\cmc)] \ <\ \cmc,
\label{cac} 
\end{equation}
%
where $F(x) \sim x$ for $x\ll 1$ and tends to unity for
$x \gg 1$.
This simple model demonstrates the physics 
governing the flexible polymer--surfactant system:
polymer--surfactant affinity induces attractive
correlations between monomers, which compete with the 
bare intra-chain repulsion.
The correlations become stronger as the {\em cmc} is approached, 
and {\it they are bound to win before reaching the cmc}, 
\ie $\cac<\cmc$.
The fact that the {\em cac} is lower than the {\em cmc} 
has been established by numerous experiments 
\cite{review_general1,review_general2}.
The argument $x=v/(w^2\cmc)$ in eq.~(\ref{cac}) 
determines the strength of the {\it effective}
polymer--surfactant interaction.
Two limiting cases arise:
(i) strong polymer--surfactant interaction ($x\ll 1$), 
where $\cac\ll\cmc$;
(ii) weak interaction ($x\gg 1$), 
where $\cac <\sim \cmc$.
Note that the distinction between strong and weak interaction
involves not only the bare polymer--surfactant interaction ($w$)
as compared to the surfactant--surfactant one ($1/\cmc$), 
but also the intra-chain interaction ($v$). This observation was
not emphasized sufficiently, in our opinion, in previous studies.

In the case of strong polymer--surfactant interaction,
$w^2\gg v/\cmc$,
the attraction between surfactants in the solution
is insignificant and the {\em cac} becomes independent of
the {\em cmc},
$
  \cac \sim v/w^2 \ll \cmc
$.
Practically, this corresponds to the case of a charged 
polymer (polyelectrolyte) interacting with an ionic
surfactant of the opposite charge \cite{review_Goddard}.
Due to strong electrostatic interactions,
the {\em cac} in such systems is usually found to be
orders of magnitude lower than the {\em cmc}.
In order for our assumption of polymer flexibility
to still hold, the system should contain additional 
salt so as to screen the electrostatic interactions on 
the length scale of the induced attractive
correlations.
Both $v$ and $w$ are expected to be dominated 
in this case by electrostatics,
and thus mainly depend on the polymer ionization degree,
$I$, and salt concentration, $c_{\rm s}$.

A surfactant-free polyelectrolyte solution is a 
complicated system by itself, whose behaviour as
function of $I$ and $c_{\rm s}$ is not completely
settled yet \cite{review_pe}.
Two observations, however, can be made:
(i) the monomer--monomer parameter, $v$, should have a 
stronger dependence on $I$ than the 
monomer--surfactant one (the simplest dependence would
be $v\sim I^2$ and $w \sim I$);
(ii) both $v$ and $w$ should have a similar 
decreasing dependence on $c_{\rm s}$.
Consequently, $\cac\sim v/w^2$ should increase with
$c_{\rm s}$ and, somewhat more surprisingly, 
be independent or weakly dependent on $I$.
The increase of {\em cac} with salt concentration was observed
in many experiments \cite{review_Goddard}.
A more detailed discussion of the dependence on $c_{\rm s}$
is postponed to a future paper \cite{future}.
The physical reason for the weak dependence on $I$ stems
from a competition between a polymer--surfactant effect 
and an intra-chain one.
A vanishing or slightly {\em increasing} dependence
on $I$ was observed in several systems involving
polyelectrolytes and oppositely charged surfactants with added
salt \cite{Benrraou92,Anthony96,cac_increase_I}.

In the case of weak polymer--surfactant interaction,
$w^2\ll v/\cmc$, according to eq.~(\ref{cac}),
the {\em cac} and {\em cmc} become comparable,
$
  \cac = A\cmc
$,
where $A <\sim 1$ is a constant which is not
very sensitive to changes in $v$, $w$ or $\cmc$.
Practically, this limit applies to systems where at least 
one of the species is uncharged 
\cite{pe_nonionic}.
The {\em cac} is expected to depend in this case on molecular
details, but {\it this complicated dependence should be 
mostly incorporated in the cmc}. 
In other words, changing various parameters
may lead to considerable changes in the {\em cmc}, 
yet the {\em cac} is expected to follow such changes roughly
linearly.
This simple prediction agrees with various experiments
\cite{pe_nonionic,cac_prop_cmc},
where proportionality between the {\em cac} and
{\em cmc} was observed over a wide range (up to two decades) 
of {\em cmc} values.

The treatment given above for the onset of self-assembly
yields a {\em cac} which can be described as a `shifted' $\theta$ 
collapse --- a sharp transition of polymer conformation 
occurring when the 2nd virial coefficient changes sign.
In practice, however, the binding of surfactants to flexible
polymers exhibits a steep, albeit continuous increase at 
the {\em cac}.
Due to the different range of competing interactions in
this case, the chain undergoes
a {\it partial collapse} \cite{deGennes}
into sub-units (`blobs'). The 
interaction between monomers within each blob is dominated 
by the short-range repulsion, whereas the interaction between
blobs is dominated by the attractive correlations.
Binding progresses {\it continuously} as additional blobs
form and the local monomer concentration increases.

The attractive potential induced between the monomers
due to surfactant correlations is assumed to have the general
form
$
  U(r) = -e^2\psi(r/\xi)
$,
where $e^2$ is a coupling constant, $\xi$ a 
correlation length, and $\psi(x)$ a dimensionless
function decaying fast to zero for $x>1$.
The two parameters, $e^2$ and $\xi$, are related to
our phenomenological interaction parameter, $w$.
Assuming weak correlations, $U<1$ (in units of $\kB T$),
we readily obtain for the effective excluded-volume parameter
of the chain,
$
  v_{\rm eff} = v + \int\drm\vecr U(r) = v - k_1 e^2 \xi^3
$,
where $k_1$ is a dimensionless constant.
Comparing to eq.~(\ref{veff}) we find
$
  e^2 \sim v_{\rm ps}/\xi^3.
$

Let us now consider blobs of size $\xi$, each containing
$g$ monomers.
The potential between blobs, $V(r)$, consists of a 
hard-core part, $V(r<\xi)\rightarrow \infty$,
and an attractive part,  $V(r>\xi)\sim g^2U(r)$, 
coming from the 
interaction of $g^2$ pairs of monomers.
The resulting excluded-volume parameter for the blobs
is 
$
  v_{\rm blob} = \int\drm\vecr [1 - \exp(-V(r))]
  \simeq k_2\xi^3 - k_3\xi^3\exp(k_4 g^2 e^2)
$,
where $k_2$,$k_3$,$k_4$ are dimensionless constants.
The condition for partial collapse is $v_{\rm blob}=0$,
\ie
$
  g^2 e^2 = \ln(k_2/k_3)/k_4 = \mbox{const}
$.
In addition, 
the blob size and number of segments are related by a 
certain power law,
  $\xi \sim g^\nu$ 
($\nu\simeq 0.6$ for a self-avoiding random walk).
Substituting this relation and the value of $e^2$
in the partial-collapse condition, we get
\begin{equation}
  g \sim (v/v_{\rm ps})^{1/\alpha}, \ \ \ 
  \xi \sim (v/v_{\rm ps})^{\nu/\alpha}; \ \ \ 
  \alpha \equiv 2-3\nu.
\label{g_xi}
\end{equation}
In order for these results to be consistent with
the physical picture in mind, 
$g$ should decrease with $v_{\rm ps}$
(\eg the entire chain should reduce to a single blob 
for small enough $\varphi_{\rm b}$).
The condition for self-consistency is, therefore,
$\alpha>0$, or $\nu<2/3$.
This yields a precise definition for our
requirement of polymer flexibility ---
on the length scale of surfactant correlations
the chain statistics should obey $\nu<2/3$.
(In particular, the chain should not be stretched,
having $\nu=1$.)
In polyelectrolyte solutions, for example, this
condition sets a lower bound for salt concentration,
below which the chain is too stretched on the length
scale of $\xi$ and the partial-collapse picture 
becomes invalid.

Repeating the calculation in $d$ dimensions
gives the same result as eq.~(\ref{g_xi}) 
with $\alpha=2-\nu d$.
Our self-consistency condition is similar to 
well known results for the critical behaviour of
disordered systems.
For both annealed and quenched impurities 
(Fisher renormalisation \cite{Fisher} and the Harris criterion 
\cite{Harris}, respectively)
the critical behaviour of the system is 
affected by impurities if $\nu<2/d$, \ie
$\alpha>0$ (the `cross-over exponent').
Similarly, surfactants affect the conformational behaviour
of a polymer only if $\nu<2/d$.
We note that for self-avoiding walk [$\nu\simeq 3/(d+2)$] this
condition is satisfied for $d<4$,
which is consistent with the fact that short-range
interactions such as the one induced by the surfactant
become irrelevant to polymer statistics for
$d\geq 4$.

The onset of association is expected when blobs can form,
\ie when $g$ becomes smaller than a certain number of monomers, 
$n$, corresponding to the finite range of surfactant-induced 
correlations. 
(Unlike the critical system discussed by de Gennes and Brochard 
\cite{deGennes}, it is the correlation amplitude ($e^2$), 
rather than the correlation length ($\xi$), which becomes 
large in a surfactant solution approaching the {\em cmc}.)
Using eqs.~(\ref{g_xi}), (\ref{veff}), and the function
$F(x)$, we find
\begin{equation}
  \cac = \cmc F[ n^{-\alpha} v/(w^2\cmc) ]
\label{cac_N}
\end{equation}
Comparison to eq.~(\ref{cac})
shows that the preceding, less refined analysis of the {\em cac}
applies to complete collapse ($g\sim 1$) rather than  
the onset of the actual partial collapse.
The similarity to Harris' results persists:
for small $\cac/\cmc$ we find from 
eq.~(\ref{cac_N}) $n\sim\varphi^{-1/\alpha}$,
which is analogous to Harris' result for the 
broadening of the critical point by impurities, 
$\Delta T/T_{\rm c}\sim \rho^{1/\alpha}$, where 
$\rho\ll 1$ is the concentration of impurities.
(Recall that the number of monomers, $n$, serves as a 
conjugate variable to $\Delta T/T_{\rm c}$ in the regular
polymer--magnetism analogy \cite{dG_book}.)


Another result of the partial-collapse
picture is that {\em at the cac} the polymer should obey
Gaussian statistics as function of polymerisation degree.
This prediction is still to be confirmed experimentally.
Some support can be found in light scattering and 
potentiometric experiments, reporting a surprisingly weak 
interaction between charged aggregates of ionic surfactants 
and neutral polymer \cite{no_interaction}. 
Contraction of the polymer at the {\em cac} was also
observed in several systems \cite{shrink}.

To complete the picture, a third scenario is to be 
considered. When the number or length
of hydrophobic side chains attached to a
hydrophilic backbone is large enough,
the polymer (known as {\em polysoap})
is already partially collapsed by itself and 
should not exhibit any further instability upon
addition of surfactant.
Hence, no sharp onset of binding ({\em cac}) is expected;
the binding to such a chain should progress
gradually as function of surfactant concentration. 
This indeed was observed for the interaction of
ionic surfactants with hydrophobically 
modified polyacids
\cite{Benrraou92,Anthony96,Zana_polysoap}. 
By synthesizing water-soluble 
polymers with various hydrophobic side-chain lengths
and controlling their ionization degree, a 
cross-over from the polysoap regime (defined above) to the 
polyelectrolyte regime could be seen. 
Indeed, it was accompanied by a cross-over from 
gradual to sharp, co-operative binding \cite{Benrraou92}.
We regard this experimental observation as a strong
support for our basic conjecture, associating the {\em cac}
with a conformational change.

\begin{figure}
\vbox to 7cm{\vfill
\centerline{
\epsfxsize=0.475\linewidth
\epsfysize=0.475\linewidth
\fbox{\epsffile{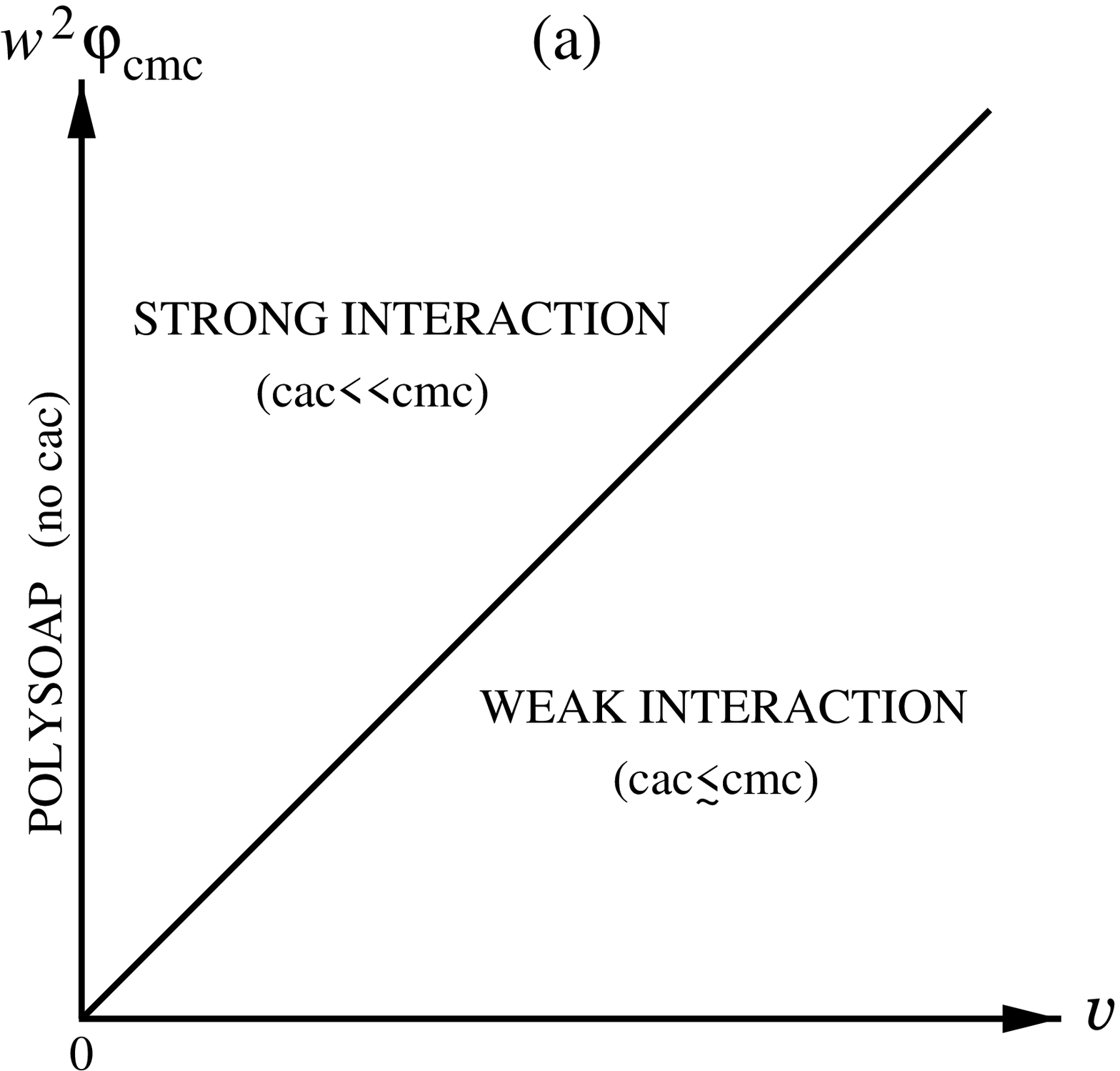}}
\epsfxsize=0.475\linewidth
\epsfysize=0.475\linewidth
\fbox{\epsffile{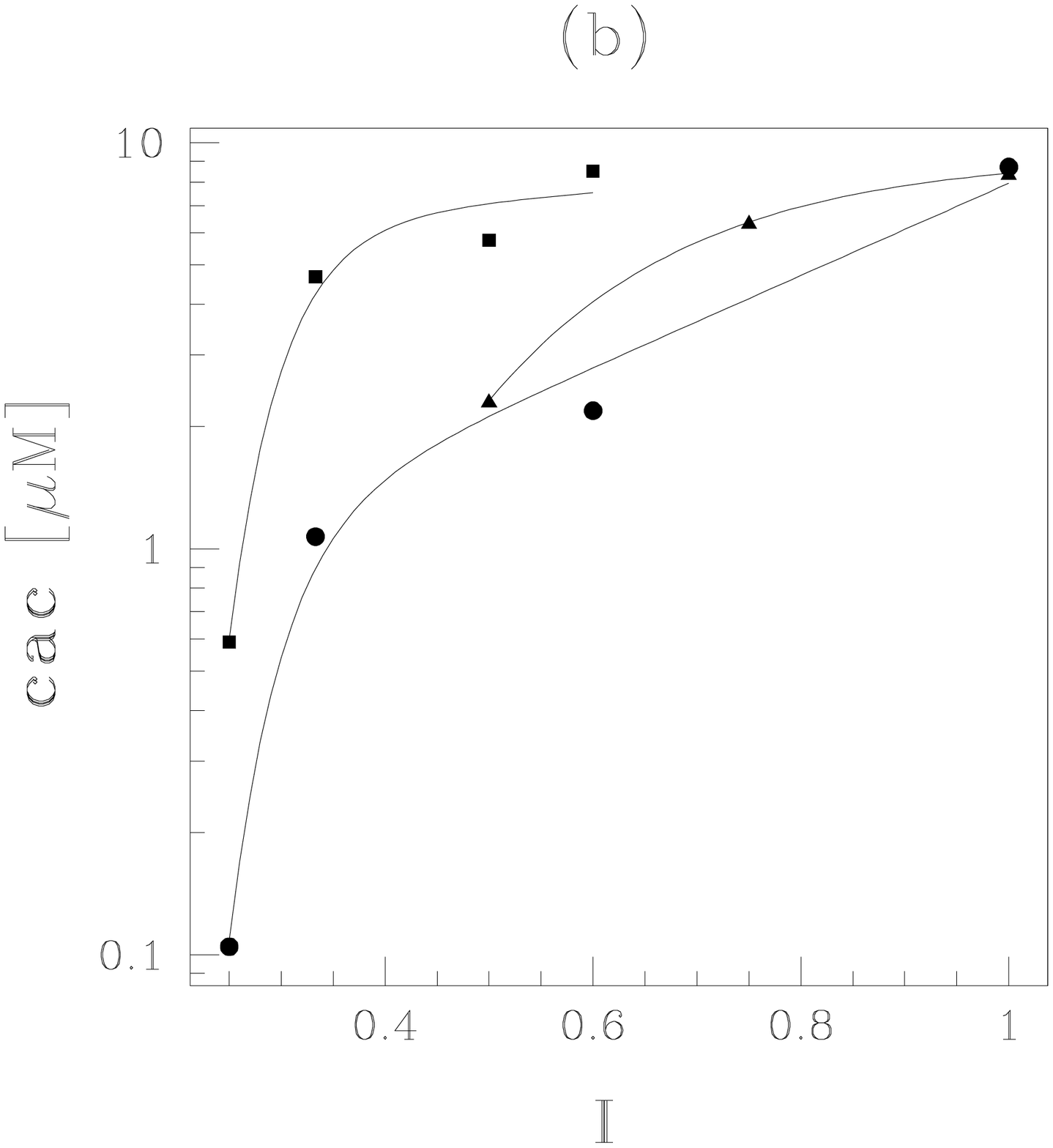}}}
\vfill}
\caption[]{(a) Summary of self-assembly regimes:
(i) a strong-interaction regime ($w^2\cmc>v$)
where $\cac\ll\cmc$, 
corresponding experimentally to systems containing
a polyelectrolyte and an oppositely charged ionic surfactant;
(ii) a weak-interaction regime ($w^2\cmc<v$) where 
$\cac<\sim\cmc$, corresponding to 
systems where one of the species is uncharged;
(iii) a polysoap regime ($v=0$) where there is gradual
binding (no {\em cac}), corresponding to polymers which form 
aggregates in the absence of surfactant.
(b) Dependence of {\em cac} on polymer charge close to 
the polysoap regime.
Triangles -- poly(maleic acid-co-butylvinylether), DTAB,
5 mM KBr (ref.~\cite{Benrraou92}).
For $I<0.5$ this polymer becomes a polysoap.
Circles -- (CH$_2$)$_x$(CH$_2$)$_y$-ionine bromide,
SDS, no salt; squares -- with 20 mM NaBr 
(ref.~\cite{Chen98}).
A distance of 3 hydrocarbon groups between charged 
groups along the backbone has been defined as $I=1$.
The lines are merely guides to the eye.}
\label{fig_diag}
\end{figure}

The diagram in fig.~\ref{fig_diag}a summarises the three 
self-assembly regimes.
Note that a cross-over to the polysoap behaviour 
($v=0$) is possible 
only from the strong-interaction regime ($w^2\cmc \gg v$).
The physical reason is that close to the
polysoap regime the stability of the 
polymer is only marginal, and a small amount of
surfactant is sufficient to cause partial collapse.
Hence, in this region of $v >\sim 0$
intra-chain features, rather than the affinity between
the two species, determine the onset of self-assembly.
As a result, the {\em cac} can be significantly reduced without 
a significant change in the bare polymer--surfactant interaction,
or, moreover, even if the bare affinity
becomes {\it weaker}.
There are two available experimental works which demonstrate
this surprising result \cite{Benrraou92,Chen98}, 
as shown in fig.~\ref{fig_diag}b.
%
%

In conclusion, focussing on the onset of self-assembly 
(the {\em cac}), we have presented a unified description of 
the interaction between a flexible polymer and surfactant in 
dilute solution.
Apart from the bare interaction between the two species,
intra-chain interactions (\ie the effective excluded-volume
parameter of the chain) are shown to have an important
role.
Utilising a conjecture of partial collapse
of the polymer at the onset of binding, simple predictions 
can be made, which seem to be well supported by experiments.
We have pointed out an interesting analogy between
the partial collapse induced by the surfactant and the
smoothing of critical behaviour by impurities 
in disordered systems.

Three self-assembly regimes are found, as shown 
in fig.~\ref{fig_diag}a. 
By modifying the polymer one can observe a cross-over
between the regimes without necessarily changing
the bare polymer--surfactant interaction.
An interesting experiment would be to take a weakly 
interacting system (\eg a polyacid like PAA and 
a nonionic surfactant such as C$_{n}$E$_m$) and by
carefully modifying the polymer gradually shift it
to the strong-interaction regime and finally to the polysoap
regime:
the {\em cac} is predicted to decrease from a value close to the
{\em cmc} to much lower values and finally to disappear.

\stars
We greatly benefited from conversations and correspondence
with T. Garel, B. Harris, I. Iliopoulos, B. Lindman,
L. Piculell, T. Witten and R. Zana. 
One of us (DA) would like to thank
L. Leibler for introducing him to 
the subject of associating polymers and for many illuminating
discussions. 
Partial support from the Israel Science Foundation founded by 
the Israel Academy of Sciences and Humanities --- Centers of 
Excellence Program is gratefully acknowledged.
HD would like to thank the Clore Foundation for financial
support.


\end{document}

%% file: euromacr.tex
\def\etal{{\hbox{{\tenit\ et al.\/}\tenrm :\ }}}

\def\And{{\rm and\ }}

\def\drm{{\rm d}}

\def\stars{\bigskip\centerline{***}\medskip}

\newif\ifboo \boofalse

\def\Review#1{\boofalse{\it #1},}
\def\Name#1{{\sc #1},}
\def\Vol#1{\ifboo Vol. {\bf #1}\else{\bf #1}\fi}
\def\Year#1{\ifboo #1\else(#1)\fi}
\def\Book#1{\bootrue{\it #1},}
\def\Page#1{\ifboo {\rm p. #1}\else{\rm #1}\fi}